# Same currents, different field: ΔF = 0 carrier-only stimulation is not temporal interference without the beat

Ido Haber[1,2,*], Giulio Tononi[2], Larissa Albantakis[2,*]

[1] Department of Biomedical Engineering, University of Wisconsin-Madison, Madison, WI, USA
[2] Department of Psychiatry, University of Wisconsin-Madison, Madison, WI, USA
* *Corresponding author: ihaber@wisc.edu, albantakis@wisc.edu*

## Abstract

Carrier-only stimulation (Δf = 0) is generally considered the most rigorous control for temporal interference (TI) stimulation, assumed to reproduce TI's kilohertz carrier field without its low-frequency beat. We show that this assumption is incorrect: without the frequency difference, the carrier fields interfere coherently, changing the peak strength, power, and direction of the resulting field depending on electrode polarity and montage geometry. Using computational field modeling, we demonstrate that these mismatches can be substantial and widespread. We therefore recommend that TI studies report electrode polarity, model both carrier-field exposures, and choose controls according to their study objectives.

Temporal interference (TI) stimulation rests on the premise that the low-frequency amplitude modulation (AM) of two kilohertz carrier fields acts on neurons beyond the carriers themselves, and there is accumulating evidence that it does (Grossman et al. 2017; Violante et al. 2023). The carriers were initially assumed to be physiologically inert (Grossman et al. 2017), but, as Peterchev pointed out, subthreshold kilohertz fields can have effects of their own (Peterchev 2025). Studies from Grossman et al. (Grossman et al. 2017) to ours (Schaeffer et al. 2026; Bruno et al. 2026) have included "carrier-only" stimulation as a control, delivering the same carrier currents without the frequency difference (ΔF = 0), and this has been considered the most rigorous comparison (Peterchev 2025; Demchenko et al. 2025). However, matching the carrier currents does not match the kilohertz field underlying the AM. Without the frequency difference, the two carrier fields interfere coherently, and the field's maxima and minima can differ substantially from those under TI.

Formally, let $E_1$ and $E_2$ be the field amplitude vectors of each electrode pair at a given point, which TI and carrier-only stimulation share in the quasi-static regime (Wang et al. 2024). In TI, the relative phase between carriers sweeps through every value once per beat, so the field necessarily reaches $HF_{max}^{TI} = \max(|E_1 + E_2|, |E_1 - E_2|)$, labeled "worst-case carrier exposure" in (Cassarà et al. 2025). However, without a carrier-frequency offset, the relative phase is fixed. With equal carriers in phase, peak $HF_{max}^{0}$ (0 for ΔF = 0) is $|E_1 + E_2|$, which matches $HF_{max}^{TI}$ where the fields are at most 90° apart but falls below it at larger angles, and to zero for equal and opposite fields (Fig. 1A,B). This also means the polarity of each carrier pair matters (Rampersad et al. 2019). Reversing one pair, which leaves TI unchanged, swaps $|E_1 + E_2|$ and $|E_1 - E_2|$ and moves the mismatch to where the fields are less than 90° apart, typically including the target. Because electrode order does not matter for TI, carrier polarity is rarely reported and may even have varied across subjects, exposing one group to two different carrier fields.

The time-averaged power, which drives heating, also differs. TI's carriers add incoherently, $P^{TI} \propto |E_1|^2 + |E_2|^2$, whereas those of carrier-only stimulation add coherently, $P^0 \propto |E_1 \pm E_2|^2$ (Cassarà et al. 2025). Where carrier-only stimulation matches TI's peak, it holds that peak continuously and deposits up to twice TI's power (for equal, parallel $E_1$ and $E_2$); where the fields interfere destructively, it deposits much less (Fig. 1B). The carrier-only field also oscillates along one fixed axis, whereas TI's peak direction swings over every beat (Fig. 1A), which matters for orientation-dependent effects such as conduction block. TI and carrier-only stimulation thus share peak and power only where $E_1$ and $E_2$ are orthogonal.

This mismatch is not an edge case, and its extent depends on the montage. Using the leadfield of one head model (Ernie), we computed $HF_{max}^{TI}$ and $HF_{max}^{0}$ for all 3,848,925 four-electrode montages on a 10–10 system, counting grey matter as mismatched where $HF_{max}^{0}$ fell more than 10% below $HF_{max}^{TI}$. The median montage left 13% of grey matter mismatched with its more closely matching polarity and 73% with the other (Fig. 1C). The further the angle between the two electrode pairs was from 90°, the more one polarity matched and the other missed (Fig. S1).

We then computed $TI_{max}$, $HF_{max}^{TI}$ and $HF_{max}^{0}$ for two published montages in 15 MRI-based head models (1 mA per pair) with TI-Toolbox (Haber et al. 2026) and averaged the maps in MNI space (Table S1). For the hippocampal montage of Violante et al. (Violante et al. 2023), carrier-only stimulation matched TI's peak in the left hippocampus at 1.95 times TI's power and left only 9% of grey matter mismatched (Fig. 1D). For the montage of Schaeffer et al. (Schaeffer et al. 2026), whose carrier-only group received 15 kHz carriers during sleep, it also matched TI's peak in the left insula, at 1.56 times TI's power, but left 51% of grey matter mismatched (Fig. 1E). Yet in the study's larger sample, the carrier peak did not differ between the TI and carrier-only groups in the frontal regions where it predicted the slow-wave response (Haber et al., in preparation). These examples show that the same currents can come close to TI's carrier field under one montage and miss it in half of grey matter under another.

Carrier-only stimulation may remain a useful comparison, but it is a different stimulus—not TI with the beat removed. Scaling its current cannot remove the mismatch, because a common factor leaves its spatial pattern unchanged, and where the fields are not orthogonal no single factor matches both peak and power. Nevertheless, the dependence on polarity can even be put to use. For example, reversing pair 2 in the hippocampal montage largely cancels the carrier field at the target (Fig. 1D), providing a target-sparing control with the same currents. We suggest that studies using carrier-only stimulation (i) report the phase lock and the electrode polarity of every channel, without which its field is undefined; and (ii) model and report $HF_{max}^{TI}$, $HF_{max}^{0}$ and both arms' carrier power, in the target and off target, alongside $TI_{max}$. Alternatively, a high-frequency beat (e.g., carriers of 4.5 and 5.5 kHz instead of 5.00 and 5.01 kHz) would preserve TI's peak field strength, total carrier power and envelope distribution, with a 1 kHz modulation that neurons likely follow far less than a slow beat, although it may not be entirely inert either.

Ultimately, no control simply separates the AM from the carrier field. Because direct kilohertz effects plausibly depend on field strength and orientation as well as frequency, a difference between TI and carrier-only stimulation may reflect the missing envelope, a lower carrier peak, a higher carrier power or a fixed field direction, and a null difference does not show that the envelope is inactive. TI studies should therefore treat carrier-only stimulation as an independently defined condition and select controls according to their objectives. Clinical studies may prioritize sensory matching with a ramp-only sham, whereas studies of TI mechanisms need controls that account for carrier-field exposure and should be planned with an explicit field model.

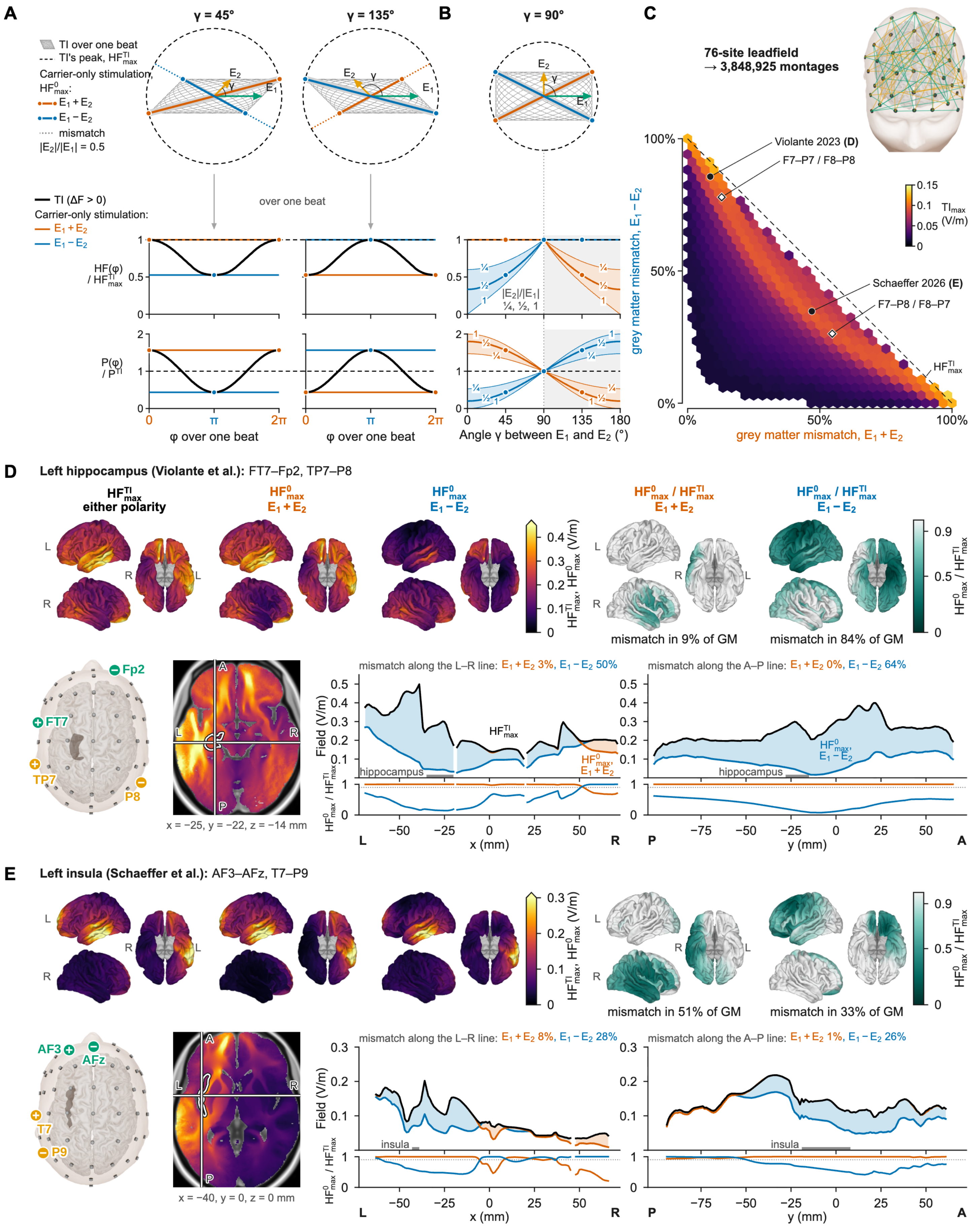

A
γ = 45°
γ = 135°
B
γ = 90°
C
TI over one beat
TI's peak, HF^TI_max
Carrier-only stimulation, HF^0_max:
E1 + E2
E1 − E2
mismatch
|E2|/|E1| = 0.5
TI (ΔF > 0)
Carrier-only stimulation:
E1 + E2
E1 − E2
over one beat
HF(φ) / HF^TI_max
P(φ) / P^TI
φ over one beat
Angle γ between E1 and E2 (°)
|E2|/|E1| ¼, ½, 1
76-site leadfield
→ 3,848,925 montages
Violante 2023 (D)
F7–P7 / F8–P8
Schaeffer 2026 (E)
F7–P8 / F8–P7
HF^TI_max
TI_max (V/m)
grey matter mismatch, E1 − E2
grey matter mismatch, E1 + E2
D
Left hippocampus (Violante et al.): FT7–Fp2, TP7–P8
HF^TI_max either polarity
HF^0_max E1 + E2
HF^0_max E1 − E2
HF^0_max / HF^TI_max E1 + E2
HF^0_max / HF^TI_max E1 − E2
mismatch in 9% of GM
mismatch in 84% of GM
Fp2
FT7
TP7
P8
x = −25, y = −22, z = −14 mm
mismatch along the L–R line: E1 + E2 3%, E1 − E2 50%
mismatch along the A–P line: E1 + E2 0%, E1 − E2 64%
Field (V/m)
hippocampus
x (mm)
y (mm)
E
Left insula (Schaeffer et al.): AF3–AFz, T7–P9
mismatch in 51% of GM
mismatch in 33% of GM
AF3
AFz
T7
P9
x = −40, y = 0, z = 0 mm
mismatch along the L–R line: E1 + E2 8%, E1 − E2 28%
mismatch along the A–P line: E1 + E2 1%, E1 − E2 26%
insula

**Fig. 1.** *At the same currents, carrier-only stimulation matches TI's peak carrier field only in regions where the two carrier fields add constructively, which are determined by the montage and electrode polarity. (A) Carrier fields of the two electrode pairs, $E_1$ and $E_2$ (first-listed electrode of each pair positive), at angles $\gamma = 45°$ and $135°$. Over one TI beat cycle, the resultant field traces the grey path and reaches $HF_{max}^{TI}$ (dashed circle). Carrier-only stimulation ($\Delta F = 0$) holds one diagonal fixed, $E_1 + E_2$, or $E_1 - E_2$ when one pair's polarity is reversed. Below each, the peak carrier field $HF(\varphi)$ and its power $P(\varphi)$ as a function of the carriers' phase difference $\varphi$ over one beat: TI sweeps through $\varphi$, reaching $HF_{max}^{TI}$, while carrier-only stimulation holds $\varphi = 0$ ($E_1 + E_2$) or $\varphi = \pi$ ($E_1 - E_2$). (B) At $\gamma = 90°$ both polarities match TI. Bottom: each polarity's $HF_{max}^{0}/HF_{max}^{TI}$ and $P^0/P^{TI}$ against $\gamma$ (lines, $|E_2|/|E_1| = ½$; bands, ¼ to 1, with edges labelled). (C) All four-electrode montages from ernie's 76-site leadfield (top right: the sites, with 40 montages drawn at random): the fraction of grey matter mismatched ($HF_{max}^{0} < 0.9\ HF_{max}^{TI}$) for $E_1 + E_2$ and $E_1 - E_2$, colored by the montages' median $TI_{max}$; circles indicate published example montages. (D, E) Means across 15 head models shown on the cortex and along two lines of an axial slice through the target ($HF_{max}^{TI}$, target outlined), with the ratio $HF_{max}^{0}/HF_{max}^{TI}$ below (dotted line, 0.9 mismatch threshold); shading indicates the mismatch (reported above each plot as a fraction of the area under $HF_{max}^{TI}$). Montages are shown on ernie, viewed from above as in the axial slice (anterior up, subject's left on the left).*

## Funding

Defense Advanced Research Projects Agency (HR00112320033); National Institute of Mental Health (1R21MH141540-01).

## Competing Interests

I.H. and G.T. are authors of TI-Toolbox, the open-source software used for the field metrics. G.T. is Chair of the Board and has a financial interest in Intrinsic Powers, Inc. L.A. has received institutional research funding from a WARF Accelerator/Nexus NeuroTech Neurotechnology Challenge Grant and an Alkermes Pathways Research Award (neither providing personal income).

## Data Availability

Code that reproduces the analyses is available at https://github.com/idossha/carrier-only-stim. It uses TI-Toolbox (https://github.com/idossha/TI-Toolbox) for the leadfield, the simulations and the field metrics.

## Author Contributions

I.H.: Conceptualization, Data curation, Formal analysis, Investigation, Methodology, Software, Visualization, Writing – original draft, Writing – review & editing. G.T.: Funding acquisition, Resources, Writing – review & editing. L.A.: Conceptualization, Formal analysis, Funding acquisition, Investigation, Validation, Visualization, Writing – original draft, Writing – review & editing.

## Declaration of generative AI and AI-assisted technologies in the manuscript preparation process

During the preparation of this work, the authors used Claude Opus 5.5 (Anthropic) in order to assist with text editing, restructuring, and clarity improvements. No scientific content was generated by AI. After using this tool/service, the authors reviewed and edited the content as needed and take full responsibility for the content of the published article.

# Supplementary Information: Same currents, different field: ΔF = 0 carrier-only stimulation is not temporal interference without the beat

Ido Haber[1,2,*], Giulio Tononi[2], Larissa Albantakis[2,*]

[1] Department of Biomedical Engineering, University of Wisconsin-Madison, Madison, WI, USA
[2] Department of Psychiatry, University of Wisconsin-Madison, Madison, WI, USA
* *Corresponding author: ihaber@wisc.edu, albantakis@wisc.edu*

## Contents

## The angle between the electrode pairs

Which polarity is mismatched follows, roughly, the angle θ between the two pairs on the scalp, each drawn + → −, which ranges from 1° to 99° across the four montages of Fig. 1 (Fig. S1A). Across all 3,848,925 montages of Fig. 1C, the share of grey matter mismatched with $E_1 + E_2$ rises with θ (Spearman $\rho = 0.92$), and its median is 7%, 40% and 84% at θ = 45°, 90° and 135° (Fig. S1B). The spread around that median is wide, so θ indicates a montage's mismatch but does not set it.

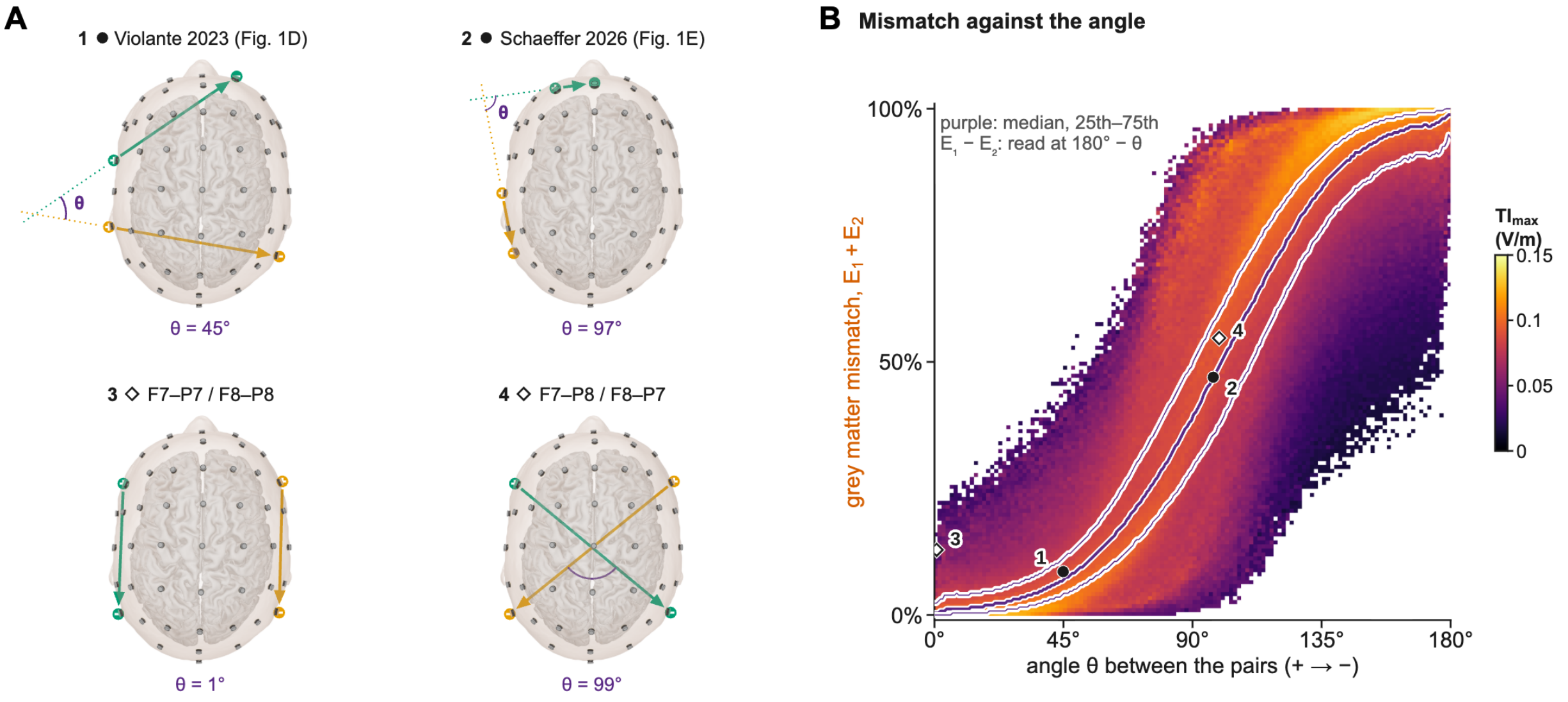

**Fig. S1.** *The percentage of mismatch between the TI and carrier-only fields is associated with the angle θ between the electrode pairs (A) How θ is measured, on the four montages of Fig. 1: those of Violante et al. (Fig. 1D) and Schaeffer et al. (Fig. 1E), and the two example montages of Fig. 1C, on ernie viewed from above. Each pair is drawn + → − between its electrodes (pair 1 green, pair 2 orange); θ is the angle between the two directions in 3D, marked where the pairs' lines meet (dotted where they are extended). (B) All 3,848,925 four-electrode montages of ernie's 76-site leadfield, each drawn for both polarities: the fraction of grey matter mismatched ($HF_{max}^0 < 0.9\, HF_{max}^{TI}$) for $E_1 + E_2$ against θ, colored by the montages' median $TI_{max}$. Lines indicate the median at each θ and its 25th and 75th percentiles, and the four montages are numbered as in A (filled circles, Fig. 1D and 1E; diamonds, additional examples of Fig. 1C).*

## The two published montages

Table S1 gives the medians and ranges over the 15 head models behind Fig. 1D and Fig. 1E. In both targets, carrier-only stimulation matched TI's peak in every head model, at about 1.5 to 2 times TI's power, while the share of grey matter mismatched was 9% for one montage and 51% for the other. The montage landscape of Fig. 1C covers every four-electrode montage of 76 sites of the 10–10 system on the leadfield of Ernie, the SimNIBS example head model.

**Table S1.** *Modeled fields of the two published montages in 15 head models.*

| | Hippocampal montage (1) | Insula montage (2) |
|---|---|---|
| Electrode pairs | FT7–Fp2, TP7–P8 | AF3–AFz, T7–P9[a] |
| Target | Left hippocampus | Left insula |
| $TI_{max}$ in target (V/m) | 0.217 (0.183–0.260) | 0.075 (0.053–0.084) |
| $HF_{max}^{TI}$ in target (V/m) | 0.248 (0.199–0.287) | 0.129 (0.105–0.148) |
| $HF_{max}^0/HF_{max}^{TI}$ in target | 1.00 (1.00–1.00) | 1.00 (1.00–1.00) |
| $P^0/P^{TI}$ in target | 1.95 (1.92–1.97) | 1.56 (1.45–1.68) |
| Grey matter mismatched | 9% (7–11%) | 51% (43–56%) |

*Note. Median (range) over 15 MRI-based head models, 1 mA per pair. Each head model contributes the volume-weighted median of each quantity in the target. Grey matter mismatched is the share of grey-matter volume where $HF_{max}^0$ falls more than 10% below $HF_{max}^{TI}$. Carrier-only values are for phase-locked channels in phase, with pair 2 as each study lists it.* [a] *The nearest 10–10 sites to the EGI electrodes reported by Schaeffer et al.*

## Supplementary References